\documentclass[conference]{IEEEtran}

\usepackage{cite}
\usepackage{graphicx}
\usepackage{amsmath,amssymb,amsfonts}
\usepackage{algorithm}
\usepackage{algorithmic}
\usepackage{textcomp}
\usepackage{xcolor}
\usepackage{hyperref}
\usepackage{listings}
\usepackage[utf8]{inputenc}

\usepackage{tikz}
\usetikzlibrary{shapes.geometric, arrows.meta, positioning}
\tikzset{
	block/.style = {rectangle, draw, text width=8em, text centered, minimum height=2em},
	line/.style = {draw, -Stealth, thick},
	hash/.style = {circle, draw, minimum size=2em, text centered},
	db/.style = {cylinder, shape border rotate=90, aspect=0.3, draw}
}

\def\BibTeX{{\rm B\kern-.05em{\sc i\kern-.025em b}\kern-.08em\TeX}}

\begin{document}
	
	\title{Modern Hardware Security:\\ A Review of Attacks and Countermeasures}
	
	\author{
		\IEEEauthorblockN{Jyotiprakash Mishra\IEEEauthorrefmark{1}\IEEEauthorrefmark{2}, Sanjay K.\ Sahay\IEEEauthorrefmark{1}}
		\IEEEauthorblockA{\IEEEauthorrefmark{1}Department of Computer Science \& Information Systems,
			BITS Pilani K K Birla Goa Campus, Zuarinagar, Goa, 403726, India \\
			Email: \{p20210041, ssahay\}@goa.bits-pilani.ac.in}
		\IEEEauthorblockA{\IEEEauthorrefmark{2}School of Computer Engineering,
			KIIT University, Bhubaneswar, Odisha, 751024, India \\
			Email: jyotiprakash.mishrafcs@kiit.ac.in}
	}
	
	\maketitle
	
	\begin{abstract}
		With the exponential rise in the use of cloud services, smart devices, and IoT devices, advanced cyber attacks have become increasingly sophisticated and ubiquitous. Furthermore, the rapid evolution of computing architectures and memory technologies has created an urgent need to understand and address hardware security vulnerabilities. In this paper, we review the current state of vulnerabilities and mitigation strategies in contemporary computing systems. We discuss cache side-channel attacks (including Spectre and Meltdown), power side-channel attacks (such as Simple Power Analysis, Differential Power Analysis, Correlation Power Analysis, and Template Attacks), and advanced techniques like Voltage Glitching and Electromagnetic Analysis to help understand and build robust cybersecurity defense systems and guide further research. We also examine memory encryption, focusing on confidentiality, granularity, key management, masking, and re-keying strategies. Additionally, we cover Cryptographic Instruction Set Architectures, Secure Boot, Root of Trust mechanisms, Physical Unclonable Functions, and hardware fault injection techniques. The paper concludes with an analysis of the RISC-V architecture's unique security challenges. The comprehensive analysis presented in this paper is essential for building resilient hardware security solutions that can protect against both current and emerging threats in an increasingly challenging security landscape.
	\end{abstract}
	
	
	\section{Introduction}
Modern hardware security faces increasing threats from a range of advanced, complex attacks that exploit both hardware and software vulnerabilities. These include side-channel attacks, Differential Power Analysis (DPA), and Electromagnetic Fault Injection (EMFI), which leverage the physical characteristics of hardware to extract sensitive information. For instance, DPA exploits variations in power consumption to deduce cryptographic keys, while EMFI induces faults to manipulate cryptographic systems. Furthermore, speculative execution attacks like Spectre and Meltdown present significant risks by leveraging speculative execution to leak sensitive data through microarchitectural side channels. These diverse and sophisticated attacks expose a wide array of hardware vulnerabilities \cite{richterbrockmann-2020-asap, fanjas-2023-dft, oflynn-2018-securecan}. In this context, the open-source, flexible, and customizable nature of the Reduced Instruction Set Computer V (RISC-V) architecture has driven its rapid adoption, fostering innovation in processor design across applications ranging from Internet of Things (IoT) devices to high-performance computing. However, this openness also introduces vulnerabilities. Recent studies have shown that the RISC-V architecture is susceptible to various microarchitectural threats, such as cache-based side-channel attacks and speculative execution flaws, which are particularly concerning in multi-tenant environments or systems where processes share resources. These critical security gaps demand robust countermeasures \cite{cassano-2022-dft, mulder-2019-dac, jin-2022-acmtaco, nashimoto-2021-tches, le-2021-ieeeaccess}.

\par To counter such threats, various hardware and software mitigation techniques have been proposed over time. Techniques such as power analysis resistance, fault detection mechanisms, and secure boot processes are now essential for ensuring the integrity and confidentiality of critical data. For example, memory encryption and secure boot processes protect systems from unauthorized access and boot-time attacks, while cache partitioning and speculative execution barriers help mitigate microarchitectural side-channel threats. Addressing these issues adequately during the design phase can significantly enhance the resilience of modern hardware systems, making them more capable of defending against sophisticated attacks \cite{meng-2023-iccd, hajiabadi-2023-arxiv, roth-2005-isca, kim-2019-ieee}. Accordingly, in Section II, we introduce the general concepts underpinning a wide range of hardware security attacks, including side-channel attacks, secure enclaves, cryptographic Instruction Set Architectures (ISAs), hardware fault injections, memory encryption, secure boot, root of trust, and Physical Unclonable Functions (PUFs). Thereafter, in Section III we review the literature on these attack vectors, combining research on both artificially created vulnerabilities and real-world exploits, alongside mitigation techniques. Due to the rising prominence of RISC-V, Section IV specifically addresses the new security challenges and countermeasures for the RISC-V architecture, discussing its open-source nature and the security concerns arising from its extensibility. Finally, Section V contains the conclusion of the paper, and provides an essential understanding of the various threats and defenses in hardware design, ensuring that future innovations are protected from emerging attack vectors.

\section{Preliminaries} 

\subsection{Cache Side-Channels}

\subsubsection{Overview of Cache Memory}
Cache memory is a small, high-speed buffer used to temporarily hold frequently accessed portions of main memory, reducing the Central Processing Unit's (CPU) data access time. Its operation is based on the principle of locality, divided into \textit{temporal locality} (recently accessed data is likely to be reused soon) and \textit{spatial locality} (data close to recently accessed memory is expected to be accessed shortly). This architecture significantly enhances system performance by lowering latency and alleviating the load on primary memory, enabling the CPU to process instructions more efficiently without delays in data retrieval \cite{Smith-1982-ACM}. Modern processors incorporate multiple levels of cache, each balancing speed, size, and proximity. The L1 cache, the smallest and fastest, resides closest to the CPU cores and is typically divided into instruction (L1i) and data (L1d) caches. The L2 cache is larger but slower and acts as an intermediary between the L1 cache and main memory. The L3 cache, or Last Level Cache (LLC), is shared among all CPU cores and helps in bridging the speed gap between the L2 cache and main memory. This hierarchical structure ensures efficient data access, with the fastest caches holding the most urgent data, while larger caches store broader datasets closer to the processor. Consistency protocols ensure that all cache levels reflect the most up-to-date data, maintaining operational integrity and preventing bottlenecks in multi-core systems \cite{Kosmidis-2013-IEEE, Paikaray-2013-ArXiv}.

\subsubsection{Side-channel Attacks}
Side-channel attacks exploit unintended information leakage from systems to extract sensitive data or gain unauthorized access. Unlike traditional attacks targeting algorithms or protocols, side-channel attacks exploit physical properties and behaviors that inadvertently reveal internal system operations. These attacks can be passive (the attacker snoops the leaked information) or active (the attacker induces faults in the system to observe the effects). Side-channel data can be used to infer confidential information, manipulate system behavior, or compromise data integrity, making these attacks a significant threat across a wide range of applications \cite{Kocher-1999-CRYPTO, Gandolfi-2001-CHES}.

\subsubsection{Types of Cache Side-Channel Attacks}
Cache side-channel attacks exploit variations in cache behavior to extract sensitive data, particularly cryptographic keys. \textbf{Time-driven attacks} exploit timing differences between cache hits and misses to infer the data accessed by a victim process. For instance, when data is fetched from the cache (cache hit), access times are significantly shorter compared to when data is fetched from lower levels of memory (cache miss). These timing differences can reveal memory access patterns, particularly in cryptographic algorithms like Advanced Encryption Standard (AES), enabling attackers to deduce secret keys \cite{Bernstein-2005-Preprint, Tromer-2009-JCryptol}. \textbf{Trace-driven attacks} use detailed traces of cache behavior to recover secret information by analyzing access patterns. By monitoring cache accesses during cryptographic operations, such as substitution-box (S-box) lookups in AES, and correlating these observations with known plaintext or ciphertext, attackers can efficiently recover secret keys. Advanced statistical and algebraic techniques can reduce the number of traces required for a successful attack, making trace-driven attacks highly effective \cite{Wan-2018-IJES, Ashokkumar-2019-Sadhana, Esfahani-2020-IACR}. \textbf{Access-driven attacks} monitor which cache lines are accessed or evicted, particularly in shared memory systems, to infer the memory operations of a victim process. In cryptographic implementations like AES, access-driven attacks can track cache lines used during T-table lookups, enabling attackers to deduce secret keys through pattern matching and statistical analysis of cache evictions and accesses. These attacks emphasize the importance of secure cache design and robust countermeasures in cryptographic systems \cite{Neve-2006-Advances, Gullasch-2011-SP, Ashokkumar-2016-EuroSP}.

\subsubsection{Cache Attack Techniques}
Cache side-channel attacks exploit timing differences in cache behavior to infer sensitive information from victim processes. The \textbf{Prime+Probe} attack involves the attacker filling the cache with their own data (priming), allowing the victim process to execute, and then re-accessing the cache (probing) to detect longer access times, revealing memory access patterns, particularly in cryptographic operations like AES \cite{Liu-2015-SP, Mushtaq-2018-GIIS}. The \textbf{Flush+Reload} attack involves flushing a memory location, allowing the victim to access it, and then reloading it to measure access times, identifying whether the victim accessed the flushed memory \cite{Yarom-2014-IACR, Gruss-2016-Flush-Reload}. \textbf{Evict+Reload} is an access-driven attack where the attacker evicts specific cache lines, allows the victim process to execute, and measures access times to infer if the evicted cache lines were accessed \cite{Liu-2015-SP, Aciicmez-2007}. \textbf{Prime+Abort} extends Prime+Probe by leveraging hardware transactional memory, such as Intel Transactional Synchronization Extensions (Intel TSX). The attacker primes the cache and initiates a transaction, and if the transaction aborts, it signals a cache conflict, revealing sensitive cache activity with high resolution \cite{Kim-2021-ICOIN, Kim-2024-TIFS}. \textbf{Flush+Flush} measures the time it takes to flush a cache line twice, where a faster second flush indicates the victim did not access the cache line. This method is less noisy and more suitable for precise timing environments \cite{Gruss-2015-ArXiv}.

\subsubsection{Speculative Execution}
Speculative execution is a performance optimization method used by modern processors to predict and execute future instructions in advance. This technique allows processors to work on instructions that may be needed shortly, thus maximizing resource utilization and minimizing idle time. If the prediction is accurate, the speculative execution results are committed, leading to a faster execution time. However, when mispredictions occur, the speculative results are discarded, and the correct execution path must be refetched, resulting in performance penalties. Nevertheless, speculative execution provides a significant boost in efficiency by overlapping latencies and keeping resources engaged \cite{Hennessy-2017-Book}. A key component of speculative execution is the \textbf{branch predictor}, which predicts the direction of branch instructions, allowing the processor to pre-fetch and execute appropriate instructions. Static, dynamic, and hybrid predictors enhance prediction accuracy by storing branch history, thus reducing pipeline flushes and improving processor performance \cite{Hennessy-2017-Book, Jimenez-2001-ISCA}. Another technique, \textbf{store-to-load forwarding}, improves memory operations by forwarding data directly from a store instruction to a subsequent load instruction targeting the same memory location, bypassing the memory hierarchy. While this reduces latency and enhances throughput, careful management is required to avoid data hazards \cite{Witharana-2022-ICCAD, Hennessy-2017-Book}. The \textbf{Return Stack Buffer (RSB)} helps to predict return addresses during speculative execution by maintaining a small stack of return addresses from function calls, improving prediction accuracy and reducing mispredictions, thus enhancing overall execution efficiency \cite{Maisuradze-2018-ACM, Korshunov-2017-SpectreRSB}.

\subsubsection{Spectre}
Spectre is a cache side-channel attack that exploits speculative execution in modern processors to infer and leak sensitive information. Although speculative execution improves performance by predicting future instructions, it can also be manipulated to access sensitive data in the cache. While speculative results are discarded, the cached data remains accessible through side-channel attacks \cite{Kocher-2020-ACM, Brotzman-2021-ACM}. \textbf{Spectre Variant 1} (Bounds Check Bypass) targets branch predictors, tricking the processor into speculatively executing out-of-bounds memory accesses. Although these accesses are discarded, traces remain in the cache, allowing attackers to recover sensitive information using cache attacks like Flush+Reload \cite{Oleksenko-2018-ArXiv, Kiriansky-2018-ArXiv}. \textbf{Spectre Variant 2} (Branch Target Injection) poisons indirect branch predictors, leading the processor to speculatively execute instructions from attacker-controlled memory, leaking sensitive data through cache side-channels \cite{Milburn-2023-EuroS&P, Koruyeh-2019-SP}. \textbf{Spectre Variant 4} (Speculative Store Bypass) exploits store-to-load forwarding, where a load retrieves stale data speculatively bypassing a store. Although this load is eventually corrected, the data remains in the cache, enabling attackers to infer sensitive information using techniques like Flush+Reload \cite{chakraborty_2022, kiriansky_2018}. \textbf{NetSpectre}, a remote variant of Spectre, expands the attack by allowing speculative execution to be exploited over a network without requiring code execution on the victim’s machine. Through remote execution of the Evict+Reload cache attack, NetSpectre leaks sensitive information, albeit with lower exfiltration rates than local Spectre attacks \cite{schwarz_2018}.

\subsubsection{Meltdown}
Meltdown is a microarchitectural vulnerability that leverages speculative execution to bypass memory protection mechanisms and access privileged memory. The attack works by speculatively executing instructions that would normally be blocked by memory protection checks, temporarily allowing the processor to access restricted memory. While the speculative results are discarded, the cache stores traces of this data, which can be extracted through side-channel techniques like Flush+Reload. This allows attackers to infer sensitive information such as encryption keys and passwords by measuring access times to various memory locations \cite{lipp_2018}. \textbf{Meltdown Variant 1} targets the boundary between user and kernel space, exploiting speculative execution to bypass hardware-enforced memory protection, enabling an attacker to read kernel memory directly from user applications \cite{lipp_2018}. \textbf{Meltdown Variant 2} (MeltdownPrime) uses Prime+Probe instead of Flush+Reload, leveraging cache invalidation mechanisms to infer the contents of privileged memory without shared memory access \cite{trippel_2018}. \textbf{Meltdown Variant 3} (Foreshadow) targets the L1 cache, allowing an attacker to extract sensitive data from protected regions such as Intel Software Guard Extensions (Intel SGX) enclaves using cache-based side-channel techniques like Flush+Reload and Evict+Time \cite{vanbulck_2019}. \textbf{Meltdown Variant 4} (Microarchitectural Data Sampling, MDS) exploits speculative execution to leak data from CPU buffers, allowing attackers to extract information like cryptographic keys using side-channel techniques like Flush+Reload and Prime+Probe \cite{schwarz_2019}. \textbf{MeltdownVariant-RSB} targets the RSB to induce mispredicted returns during speculative execution, establishing a low-noise cache covert channel to leak sensitive information across different CPU architectures without relying on Intel-specific features \cite{koruyeh_2018}.

\subsection{Power Side-Channels}
Power side-channel attacks exploit the physical emanations of electronic devices to extract sensitive information such as cryptographic keys. These emanations include power consumption, electromagnetic radiation, and even acoustic noise, which can reveal insights into the internal operations of a device. The core idea behind power side-channel attacks is that these physical signals, which correlate with specific computational activities, are measured and analyzed to uncover confidential data. For example, attackers can monitor a CPU’s power consumption during cryptographic operations and use Simple Power Analysis (SPA) or DPA to deduce secret keys. Electromagnetic side-channel attacks use radiated emissions to infer processed data, while acoustic side-channel attacks analyze sound patterns produced by the device. These techniques emphasize the need for robust countermeasures to protect against side-channel vulnerabilities \cite{gandolfi_2001, mangard_2008}.

\subsubsection{Simple Power Analysis}
Simple Power Analysis (SPA) is a direct side-channel attack that exploits power consumption variations during cryptographic operations. By examining power traces, attackers can deduce which operations, such as multiplication or addition, are being executed. Cryptographic algorithms like Rivest–Shamir–Adleman (RSA) and Data Encryption Standard (DES) exhibit distinct power usage during specific operations. For instance, in RSA, power consumption patterns during modular exponentiation may reveal key bits, and in DES, power traces from S-box lookups can expose key-related information. SPA is particularly effective in scenarios where cryptographic operations exhibit clear power consumption differences, allowing attackers to reduce the key space and recover secret keys \cite{kocher_1999, messerges_1999, li_2012, datsios_2013}.

\subsubsection{Differential Power Analysis}
DPA is a more advanced side-channel attack that statistically analyzes multiple power consumption traces from cryptographic operations to extract secret information. DPA measures variations in power consumption for different inputs and uses this data to identify correlations between the power consumption and secret information. Attackers generate hypothetical key values and predict power consumption models, correlating these models with actual traces to pinpoint the correct key. DPA is especially effective against symmetric cryptographic algorithms like AES and DES, where intermediate values during encryption can leak key information. Devices without proper countermeasures, such as masking or randomization, are particularly vulnerable to DPA \cite{kocher_1999, messerges_2002, pramstaller_2004, strachacki_2008}.

\subsubsection{Correlation Power Analysis}
Correlation Power Analysis (CPA) is an advanced variation of DPA that focuses on correlating power consumption measurements with hypothetical values derived from intermediate stages of cryptographic algorithms. CPA uses statistical techniques to find correlations between actual power traces and predicted power consumption models based on guessed intermediate values, such as S-box lookups in AES. The highest correlation reveals the correct key. CPA is highly effective against cryptographic algorithms involving predictable operations like DES and AES, and it is frequently used to target devices such as smart cards or hardware security modules \cite{putra_2019, fang_2010}.

\subsubsection{Template Attacks}
Template attacks are a powerful side-channel attack method that involves creating detailed statistical models, or “templates,” based on a device’s power consumption during cryptographic operations. These templates are built during a profiling phase on a device with known keys, capturing characteristics like the mean and variance of power traces. Once constructed, attackers match power traces from the target device to the templates, enabling the extraction of secret information. Template attacks are highly effective, even when only a single trace is available from the target device, and they work against systems implementing countermeasures for CPA and DPA \cite{shanbehzadeh_2017, xu_2017, elaabid_2007}.

\subsubsection{Voltage Glitching and Fault Injection}
Voltage glitching and fault injection attacks manipulate a device’s power supply or clock signal to introduce faults, causing the device to malfunction during cryptographic operations. These faults may lead the device to skip instructions or bypass security checks, thereby exposing sensitive data. Voltage glitching, a common form of fault injection, briefly reduces the supply voltage, resulting in timing errors that reveal confidential information. Attackers first profile a similar device to identify the optimal conditions for glitching, and then apply voltage glitches to the target during cryptographic computations, leaking sensitive data such as cryptographic keys \cite{anderson_1997, tunstall_2007}.

\subsubsection{Electromagnetic Analysis}
Electromagnetic Analysis (EMA) is a non-invasive side-channel attack technique that captures electromagnetic emissions from a device during cryptographic operations to extract sensitive information. By placing an electromagnetic probe near the device, attackers can analyze the emitted signals, which correspond to the device’s internal data processing and electrical activity. EMA is particularly effective because it requires no physical tampering and can be conducted remotely. The attack typically involves two phases: profiling and attacking. During profiling, attackers build statistical models using electromagnetic traces from a reference device, and in the attacking phase, these models are compared to traces from the target device to deduce secret data like cryptographic keys \cite{quisquater_2001, agrawal_2002}.

\subsection{Memory Encryption}

\subsubsection{Confidentiality}
Confidentiality in memory encryption is crucial for safeguarding sensitive data from unauthorized access, especially in environments where physical access to hardware is possible. Encrypting data as it is written to memory and decrypting it when read ensures that any data captured from the memory hardware remains unreadable. This protection is vital against physical attacks like cold boot attacks, where attackers retrieve residual data from memory after a system shutdown \cite{halderman-2008-usenix}. Technologies such as Intel’s Total Memory Encryption (TME) and AMD’s Secure Memory Encryption (SME) offer transparent encryption at the hardware level, providing strong protection without requiring software modifications. These methods are particularly valuable in environments where physical access is a risk, as they prevent data theft even if attackers gain direct access to memory hardware \cite{mofrad-2018-proceedings, gueron-2016-ieee}.

\subsubsection{Authenticity}
Authenticity in memory encryption focuses on ensuring the integrity of stored data and defending against spoofing and replay attacks. Integrity checks, such as cryptographic hashes and integrity trees, verify that data has not been altered or tampered with, preventing unauthorized modifications. Cryptographic authentication methods ensure that data originates from a trusted source, mitigating spoofing attacks. Replay attacks, where attackers attempt to reuse previously valid data to trick the system, are countered using nonces or timestamps, ensuring that each data transaction is unique and cannot be reused maliciously \cite{shwartz-2018-ieee, elbaz-2007-lecture}.

\subsubsection{Granularity}
Granularity in memory encryption refers to the level at which encryption is applied within the memory hierarchy. Fine-grained encryption focuses on smaller units, such as cache lines, registers, or specific memory pages, allowing more precise control over what needs protection, which helps minimize performance overhead. This method is useful when only critical sections of memory require encryption. Both TME and SME employ page-level encryption, balancing security and performance. TME provides full memory encryption, including at the page level, to protect against physical attacks with minimal performance impact \cite{gueron-2016-ieee}. Similarly, SME encrypts data at the page level, ensuring protection against unauthorized access with minimal performance degradation \cite{sartakov-2019-edcc}. Coarse-grained encryption, on the other hand, encrypts larger memory regions, such as entire memory segments or swap files. While this simplifies the process, it may lead to increased performance overhead due to the larger encrypted areas.

\subsubsection{Storage of Keys}
The secure storage of cryptographic keys is essential to ensure data integrity and authenticity in encrypted memory systems. One efficient method is using Merkle trees, which securely verify large datasets using a single root hash. Each leaf node in the tree represents a hash of a data block, while internal nodes correspond to hashes of their child nodes. Any modification to a data block changes the root hash, enabling tamper detection with minimal overhead \cite{merkle-1980-springer}. Another approach involves hierarchical key structures for managing cryptographic keys. In this model, a master key generates derived keys for specific tasks or data segments, providing fine control over key management. This allows for efficient encryption at various granularities without the need to store all individual keys \cite{wallner-1997-rfc}.

\subsubsection{Masking and Rekeying}
Masking is a well-established technique for defending against power analysis attacks by concealing intermediate cryptographic computations. This is done by combining sensitive data with random values, or “masks,” before any operation, which effectively disrupts the correlation between the processed data and the device’s power consumption. The formalization of masking, introduced by Chari et al. (1999), laid the foundation for its use in reducing vulnerabilities to side-channel attacks such as DPA and Correlation Power Analysis (CPA) \cite{chari-1999-springer}. By increasing the number of power traces required for a successful attack, masking makes the attacker’s task significantly harder \cite{messerges-2002-springer, mangard-2007-springer}. Common masking techniques include Boolean masking for logical operations and arithmetic masking for numerical operations, ensuring that masked computations appear as random noise to an attacker \cite{prouff-2011-springer, kim-2010-ieee}.

Rekeying, the practice of frequently updating encryption keys, is critical to reducing the risk of long-term key exposure. Canetti, Halevi, and Katz (2003) advanced this concept by developing efficient protocols for periodic key updates to maintain robust security in symmetric encryption schemes \cite{canetti-2003-springer}. Rekeying can be combined with masking to create layered defenses. Some systems employ rekeying intervals based on the number of side-channel leakage traces required for a successful attack \cite{vuppala-2020-ieee}. This approach minimizes overhead while enhancing security, often using hierarchical key management in which a master key generates session keys, simplifying key distribution and renewal processes \cite{boneh-2001-springer}.

\subsection{Secure Enclaves}
Secure enclaves, also known as \textit{Trusted Execution Environments (TEEs)}, are specialized areas within a processor designed to securely execute code and protect sensitive data from external threats. Their primary purpose is to provide a protected execution environment that isolates critical operations and data from the rest of the system, including the operating system and other applications. The concept of secure enclaves has evolved significantly, with early implementations focused on isolating sensitive computations from software-based attacks. The development of TEEs, such as Intel’s Software Guard Extensions (SGX) and ARM’s TrustZone, has been driven by the growing need for robust security solutions in the face of increasingly sophisticated data breaches and cyber-attacks \cite{mckeen-2013-acm, kinney-2004-arm}. Secure enclaves play a vital role in modern processor architectures, providing hardware-based solutions to ensure the integrity and confidentiality of sensitive computations, particularly in cloud computing, financial services, and other security-critical domains \cite{costan-2016-cryptology}. They form the foundation for trusted systems, enabling secure remote attestation, data sealing, and other essential security functions that strengthen the security posture of modern computing environments \cite{arnautov-2016-usenix, christensen-2017-usenix}.

Secure enclaves, like Intel's SGX and ARM's TrustZone, ensure hardware-based isolation to protect sensitive computations and data from unauthorized access. These enclaves create isolated memory regions, protected by encryption mechanisms such as Intel's Memory Encryption Engine (MEE), ensuring that data remains secure even if an attacker gains physical access to the hardware \cite{mckeen-2013-acm, gueron-2016-ieee}. A minimal \textbf{Trusted Computing Base (TCB)} is crucial in enclave design to reduce the attack surface and enhance security, as demonstrated by systems like SCONE \cite{arnautov-2016-usenix}. Remote attestation protocols further strengthen enclave security by enabling external parties to verify the integrity of the enclave’s code and hardware environment, ensuring it remains uncompromised \cite{anati_2013, schuster_2015}. Memory protection techniques, including access control and encryption, help mitigate side-channel attacks, with additional defenses like \textbf{Trusted SGX (T-SGX)} providing robust protection against memory-based exploits \cite{shih_2017, azab_2014}.

\subsection{Cryptographic ISAs}
Cryptographic ISAs integrate cryptographic operations directly into the processor, significantly improving performance and security by accelerating algorithms such as AES, Secure Hash Algorithm (SHA), and RSA. Examples include Intel's AES New Instructions (AES-NI) and RISC-V’s custom instruction extensions, which reduce the computational overhead associated with cryptographic tasks, resulting in faster encryption and decryption \cite{gueron-2010, stoffelen-2019, marshall-2020}. These hardware-accelerated instructions are not only more efficient than software-based cryptography but also offer better resistance to side-channel attacks \cite{tillich-2007}. ARM's Cryptographic Extension also supports efficient encryption and hashing operations, enabling secure and fast cryptographic processing in embedded systems \cite{baker-2013}. These advancements enable developers to implement secure cryptography with minimal complexity, improving both performance and security across a wide range of applications \cite{guo-2010}.

Cryptographic ISAs have been incorporated into various processor designs to enhance cryptographic performance and security. Intel’s AES-NI accelerates AES encryption and decryption, offering significant performance improvements while also strengthening resistance against side-channel attacks \cite{gueron-2010}. ARM’s Cryptographic Extensions in the ARMv8 architecture support AES and SHA, enabling efficient cryptographic processing for mobile and embedded devices \cite{baker-2013}. RISC-V, an open-source ISA, introduced cryptographic extensions such as Zkne and Zknh to support AES and Secure Hash Algorithm-256 (SHA-256), delivering both flexibility and efficiency for cryptographic tasks in embedded systems \cite{shaolin_2020}. Even without specific cryptographic extensions, ARMv8 can use Advanced SIMD (ASIMD) instructions to boost AES performance through parallel processing \cite{albahar_2016}. Specialized processors, such as those based on IEEE 1363, accelerate elliptic curve and Galois Field arithmetic, providing low-power secure solutions \cite{chang_2007, gallais_2008, kaeslin_2019}. Additionally, lightweight cryptography extensions in RISC-V further optimize ciphers like GIFT using bitslicing techniques, making them ideal for constrained environments \cite{bansod_2017}. Finally, Field-Programmable Gate Array (FPGA)-based cryptographic coprocessors offer algorithmic agility, enabling dynamic reconfiguration of cryptographic algorithms for high-speed, efficient processing across multiple platforms \cite{li_2018}.

\subsection{Secure Boot and Root of Trust (RoT)}
Secure Boot and RoT are essential security mechanisms that ensure the integrity of the boot process in modern systems. Secure Boot ensures that only trusted software is loaded during the boot process by verifying each component in the boot chain, starting from the Basic Input/Output System (BIOS) or Unified Extensible Firmware Interface (UEFI) firmware and extending to the operating system kernel and drivers. This chain of trust prevents unauthorized code from being introduced during startup \cite{wojtczuk_2011, intel_2013, markon_2014}. The RoT, often implemented via hardware such as Trusted Platform Modules (TPMs), securely stores cryptographic keys and conducts integrity checks to support the boot process \cite{grawrock_2006, joseph_2017}. TPMs play a critical role by creating and storing cryptographic measurements in Platform Configuration Registers (PCRs), which are used to verify system integrity during startup \cite{lin_2012, wojtczuk_2011}. RoT implementations, like ARM TrustZone and AMD Secure Encrypted Virtualization (SEV), combine hardware and firmware to create secure environments from the moment the system is powered on, providing enhanced protection against various attacks \cite{kaplan_2016, zhao_2014}. Together, Secure Boot and RoT form a robust defense mechanism that ensures the trustworthiness of critical system components from the beginning of the boot process \cite{costan_2016, grawrock_2006}.

\subsection{Physical Unclonable Functions}
PUFs are hardware-based security primitives that leverage microscopic manufacturing variations in semiconductors to create unique device “fingerprints” \cite{gassend_2002, maes_2010}. These variations result in distinct responses when the device is subjected to specific input challenges, making PUFs highly secure and resistant to cloning attacks \cite{guajardo_2007}. Despite their inherent randomness, PUFs offer high reproducibility under stable environmental conditions, with error correction techniques often employed to address noise and environmental fluctuations \cite{herder_2014, maes_2013}. PUFs are unclonable, tamper-evident, and capable of generating cryptographic keys on demand, eliminating the need to store keys in non-volatile memory, which reduces the risk of key exposure \cite{pappu_2002, delvaux_2018}. Modern processors incorporate PUFs to enhance device authentication, memory protection, and secure boot processes, making them integral to secure hardware design \cite{tehranipoor_2014, devadas_2008}.

\subsection{Hardware Fault Injection}
Hardware fault injection is a critical method for testing the reliability and robustness of computer systems by deliberately introducing faults into hardware components and analyzing their effects. This technique helps identify vulnerabilities and assess fault-tolerance mechanisms. Common techniques include \textit{voltage glitching}, which disrupts the power supply to induce faults, and \textit{clock glitching}, where timing errors are created by manipulating the clock signal \cite{skorobogatov_2012, karaklajić_2013}. Other non-invasive methods, such as \textit{EMFI} and \textit{laser fault injection}, use electromagnetic emissions or focused laser beams to introduce faults \cite{schroder_2015, skorobogatov_2010}. These techniques are particularly valuable for evaluating the security of cryptographic systems by exposing hardware implementation vulnerabilities. Fault injection also helps in understanding transient faults, such as \textit{single event upsets (SEUs)} caused by high-energy particles, or \textit{multiple bit upsets (MBUs)} that affect neighboring memory cells \cite{baumann_2005, liu_2004}. Mitigation techniques, such as error correction codes and redundancy, are typically employed to address these faults \cite{shivakumar_2002}. In environments with high radiation, such as space, where both transient and permanent faults are more common, fault injection helps improve fault-tolerant designs \cite{calhoun_2008}.

\section{Attacks and Mitigation}
\subsection{Speculative Attacks}

\textbf{Spectre Variant 1} exploits speculative execution to bypass bounds checks, allowing unauthorized memory accesses that reveal sensitive data through side channels, most often the CPU cache \cite{kocher2019spectre}. The key mechanism involves training the processor’s branch predictor to mispredict a conditional branch, causing out-of-bounds reads during speculation. Since these speculatively accessed values are eventually discarded when the misprediction is discovered, the data remains hidden in architecture-visible state—yet still leaves microarchitectural traces, such as cache footprints, that can be measured. To address this attack, Retpoline modifies indirect branch usage to prevent the CPU from speculating across indirect branches, while memory fencing instructions (e.g., LFENCE) block speculation at selected points \cite{turner2018retpoline}. Additional compiler-based instrumentation can also insert serializing instructions at locations identified as high risk, thus restricting the window during which speculation can leak data.

\textbf{Spectre Variant 2} (Branch Target Injection) manipulates the branch target buffer (BTB) to redirect speculative execution toward malicious code sequences, which can then leak data via side-channel observations \cite{kocher2019spectre}. The attack proceeds by training the BTB with specific addresses, prompting the CPU to speculatively jump to those addresses if the same branch instruction is encountered again. In this scenario, unauthorized instructions execute transiently, affecting the cache state in ways observable to an attacker. Countermeasures include Indirect Branch Restricted Speculation (IBRS) and Single Thread Indirect Branch Predictors (STIBP), which limit the sharing of branch prediction resources between contexts \cite{intel2018spectre}. These mechanisms ensure that branch predictions linked to sensitive code paths are invalidated or constrained when switching privilege levels or when operating in single-threaded mode, thus reducing unintended leakage.

\textbf{Spectre Variant 4} leverages speculative execution of loads before preceding stores are fully resolved, which can allow an attacker to bypass normal memory-safety checks and read sensitive data \cite{kiriansky2018speculative}. The CPU speculates that a store to an address has completed, but if the store is delayed or redirected, the load may retrieve data from a different location. This discrepancy can be measured through microarchitectural side channels, leaking information not meant to be accessible. Mitigation commonly involves disabling speculative store bypass, which prevents the CPU from issuing a load out-of-order with respect to a prior store to the same address space \cite{intel2018spectre}. Processors may also introduce additional ordering or fencing instructions to ensure the architectural order of memory operations is respected in contexts susceptible to these transient leaks.

\textbf{SpectreRSB} focuses on the return stack buffer (RSB), which is a hardware structure used to predict return addresses for function calls \cite{koruyeh2018spectrersb}. An attacker can deliberately unbalance the call/return pattern or insert calls that do not match subsequent returns, causing the CPU to speculate with a mismatched return address. This can lead to transient execution of instructions at attacker-chosen locations, followed by cache-based side-channel leakage. Recommendations include increasing RSB capacity and implementing hardware-level return address validation to verify that return predictions match legitimate call sites \cite{intel2019rsbsize}. Software-based workarounds can also involve “RSB stuffing,” where valid returns are preloaded into the buffer upon context switches or VM exits.

\textbf{Spectre-STL} (Store-to-Load) manipulates dependencies between store and load instructions, triggering a speculative load before its corresponding store completes \cite{canella2019spectrestl}. This reordering may allow a load to retrieve stale or privileged data, subsequently leaking the information via microarchitectural side effects. Typical mitigations include disabling speculative store bypass so that dependent loads cannot reorder ahead of their preceding store instructions, along with compiler transformations that reorder instruction sequences to limit speculative windows \cite{intel2019speculative}. These adjustments help ensure that loads do not rely on uncommitted stores, reducing the time frame in which any leaked data can be accessed.

\textbf{Spectre-PHT} abuses the Pattern History Table (PHT) to skew branch predictions, allowing speculative paths to reveal privileged information \cite{chowdhury2021branchspectre}. By selectively training the PHT with misleading branch outcomes, an attacker can cause the CPU to speculatively follow a wrong path, enabling unauthorized memory accesses that leave microarchitectural traces. Defensive strategies involve enhancements to branch prediction hardware and randomization of prediction logic to lessen the predictability of branch outcomes \cite{intel2019branchprediction}. Additionally, software can introduce “boundary fences” or other serialization instructions around critical condition checks, which ensure that speculation does not proceed into unsafe regions.

\textbf{Spectre v1.1} extends the concept of Spectre Variant 1 by not only reading out-of-bounds data but also speculatively overwriting data in memory regions, possibly altering execution flow or opening further opportunities for data exfiltration \cite{kiriansky2018speculative}. This method can be particularly dangerous if speculative writes affect instructions or pointers that will be consumed by subsequent speculative paths. Suggested mitigations include turning off speculative store bypass, allowing stores to commit in strict order, and applying fences that prevent any speculation past risky instructions \cite{weisse2019nda}. By restricting out-of-bounds writes in speculation, systems reduce the possibility that critical data structures will be improperly modified.

\textbf{Spectre v1.2} exploits speculative execution in read-only memory segments, allowing speculative writes that could overwrite supposedly immutable data \cite{kiriansky2018speculative}. Once overwritten, even if only transiently, these changes may affect subsequent speculative instructions, leading to further side-channel leakage. To address this, hardware enhancements can block speculative writes to read-only pages, while operating systems may enforce stricter protections on memory regions flagged as read-only \cite{intel2019speculative}. Compiler-based or OS-level mechanisms can similarly ensure that any attempts to speculate with write permissions on read-only segments are invalidated or serialized.

\textbf{BranchScope} manipulates directional branch predictors by altering the history they use to decide whether a branch is likely to be taken or not \cite{evtyushkin2018branchscope}. The attacker runs code that forces predictable shifts in the predictor’s internal state, then observes execution timings or other side effects to infer victim process behavior. This information can subsequently leak privileged data. Countermeasures involve randomizing branch predictor state between context switches and applying hardware-based branch validation checks to limit external code’s impact on prediction logic \cite{intel2019branchprediction}. Some systems also flush branch prediction state when transitioning between different security domains.

\textbf{MicroScope} repeatedly induces speculative faults—such as page faults or exceptions—along the same execution path to enhance side-channel signal quality \cite{skarlatos2020microscope}. By forcing the CPU to restart speculation multiple times at a vulnerable instruction sequence, attackers can collect repeated measurements that improve data extraction accuracy. A technique called Delay-on-Squash introduces an intentional wait period after speculation is squashed, reducing the frequency of these retries and thereby limiting the attacker’s ability to obtain multiple high-quality samples \cite{sakalis2022delayonsquash}. Additional control-flow or memory-fencing instructions in software can complement this strategy to constrain speculative re-execution.

\textbf{SMoTherSpectre} combines speculative execution with port contention inside the CPU, creating a scenario where the attacker monitors how instructions compete for limited execution resources \cite{bhattacharyya2019smotherspectre}. By measuring the latency or throughput of these ports, it becomes feasible to deduce which instructions the victim is speculatively executing, leading to data leaks. Strategies to reduce exposure include isolating resources associated with critical code paths or placing a process on dedicated hardware threads when sensitive operations are in progress \cite{intel2019speculative}. Fine-grained scheduling and partitioning of execution resources can help contain the spread of leak-prone transient signals.

\textbf{Load Value Injection} (LVI) subverts load operations by injecting attacker-supplied values into speculative paths, tricking the CPU into using these spoofed values in subsequent transient instructions \cite{bulck2020lvi}. As speculation proceeds, it may reveal secrets or manipulate the program’s internal state to generate new side-channel vulnerabilities. Microcode updates that introduce additional fencing or sequence constraints around load operations can reduce these possibilities \cite{weisse2019nda}. Furthermore, compilers and runtime systems can instrument code to avoid depending on untrusted loads, thus constraining the CPU’s ability to speculate on potentially malicious inputs.

\textbf{Fallout} targets the store buffer, a microarchitectural structure that temporarily holds data to be written to memory \cite{lipp2019fallout}. Through carefully timed speculative loads, an attacker can retrieve data that has not yet been fully retired or that belongs to another context, thus disclosing privileged information. Mitigation techniques often include flushing or invalidating the store buffer upon context switches, ensuring that data from one process is not visible to another \cite{intel2019speculative}. Some systems also serialize loads that depend on recently stored values, decreasing the opportunity for speculation to expose stale or privileged data.

\textbf{CacheOut} takes advantage of cache eviction sequences to determine which data is evicted from the L1 cache, thereby discovering sensitive information through timing measurements \cite{schaik2020cacheout}. The attack repeatedly flushes specific cache lines to see how eviction impacts the victim’s data usage. Defenses encompass randomizing eviction policies to make these patterns less predictable and implementing secure cache partitioning, which assigns separate cache regions to different processes \cite{jaamoum2022randomization}. In some architectures, it is also possible to invalidate or flush cache lines on context switch, reducing cross-process leakage.

\textbf{ZombieLoad} exploits fill buffers, hardware structures that hold data in transit between CPU caches and the execution pipeline. By triggering loads that require partial data fills, an attacker can observe partial data from other processes or privilege levels \cite{schwarz2019zombieload}. Frequent buffer flushing is one approach, ensuring that remnants from sensitive contexts are not retained when switching to untrusted ones \cite{intel2019speculative}. Operating systems can further restrict speculation in code sections where fill buffers might handle privileged data, and microcode updates can introduce checks on data flow within fill buffers to prevent cross-domain disclosure.

\textbf{RIDL} (Rogue In-Flight Data Load) attacks unarchitected CPU buffers—those that are used internally for scheduling or reordering instructions but are not exposed as part of the official architecture \cite{vanschaik2019ridl}. By inducing contention or stalls within these buffers, an attacker can cause secret data from other processes or kernel space to appear transiently in a measurable state. Recommended approaches include clearing these internal buffers whenever switching between processes or privilege levels, reducing the window where leakage can occur \cite{intel2019speculative}. Hardware modifications can also help isolate the buffers used by different logical or physical CPU cores.

\textbf{NetSpectre} extends speculative execution attacks over a network interface, removing the need for local code execution on the victim’s machine \cite{schwarz2018netspectre}. Attackers craft network packets that trigger specific speculative paths, which in turn modify microarchitectural state (such as cache lines), allowing data to be inferred remotely through side-channel analysis. Defenses include careful packet filtering at the OS or hardware level so that suspicious network patterns do not induce speculation in sensitive contexts. In addition, many systems adopt speculation barriers around I/O routines to ensure that data from network processing cannot be speculatively misused.

\textbf{Branch Shadowing} alters the outcomes of conditional branches so that the CPU’s prediction logic steers execution into attacker-chosen paths \cite{chowdhury2020branchshadowing}. Through repeated training of the predictor, an attacker can force transient execution of instructions that handle sensitive data, ultimately yielding side-channel leaks. Defenses involve randomizing or sanitizing predictor entries upon context switches and placing speculation barriers at critical condition checks \cite{milburn2023mitigations}. By doing so, any mismatch between the trained and actual branch outcomes is resolved more promptly, limiting the window for speculative leakage.

\textbf{Meltdown} exploits the boundary between user and kernel space by leveraging out-of-order execution to read kernel memory from user mode, thereby bypassing conventional protection checks \cite{lipp2018meltdown}. The underlying mechanism relies on instructions that cause a transient read of kernel data even though architectural rules eventually forbid it; the transient read leaves a microarchitectural footprint observable through a side channel. Kernel Page-Table Isolation (KPTI) restricts the kernel address space from user processes, thereby reducing the attacker’s ability to speculatively access kernel data \cite{gruss2018kpti}. Modern hardware revisions also introduce mechanisms that block user processes from referencing kernel mappings during speculation, decreasing the effectiveness of the attack.

\textbf{Meltdown-RW} bypasses read/write privileges in protected memory regions through speculative execution, enabling an attacker to read or even manipulate data that should be off-limits \cite{canella2019meltdownrw}. Similar to the original Meltdown, the key vector is transient execution, which temporarily ignores normal privilege checks and leaves behind microarchitectural traces. The primary approach involves applying hardware and firmware updates to strengthen read/write privilege checks, ensuring that any attempt to speculatively access protected pages triggers a stall or invalidation before data is leaked. Some operating systems also add additional verification steps for read/write operations, ensuring they cannot proceed speculatively without the correct permissions being established.

\textbf{Meltdown-US} (User/Supervisor) exploits transient execution to cross the boundary between user and supervisor privilege levels, granting unauthorized access to supervisor-only memory regions \cite{lipp2018meltdown}. The speculative instruction flow bypasses the usual ring-based security checks, exposing contents that belong solely to the OS kernel. Mitigations rely on reinforcing separation between user and supervisor memory mappings, either via KPTI or related address-space isolation techniques. Hardware vendors also introduced speculative behavior changes that invalidate or block references to supervisor pages when executing in user mode, thereby reducing the risk of transient reads.

\textbf{Meltdown-BR} (Bounds Check Bypass) targets situations where bounds checks in privileged code are speculatively bypassed, enabling reading of protected memory regions outside the permitted range \cite{weisse2019meltdownbr}. The CPU’s speculation skips over these checks under certain out-of-order execution conditions, leading to transient data exposures. The primary mitigation path involves tightening bounds checks so that they force pipeline flushes or introduce memory fences whenever a check fails. In addition, compilers and runtime systems can insert more explicit constraints, preventing the CPU from speculatively passing bound verification code when it has not been conclusively validated.

\textbf{Meltdown-PK} (Protection Key Bypass) exploits the memory protection key mechanism during speculative execution to access restricted memory \cite{canella2019meltdownpk}. Under normal circumstances, protection keys ensure that certain memory areas can only be accessed by processes holding the correct key bits in a special register. However, speculative execution may transiently ignore these checks, leading to unauthorized data leaks. Mitigation focuses on more robust enforcement of protection keys, ensuring that any speculative access to a key-protected region stalls or is invalidated if the executing process lacks the correct privileges. In some cases, additional hardware checks verify that the instruction pipeline respects key-based restrictions even during transient stages.

\subsection{Patterns of Mitigation for Speculative Attacks}

\textbf{Speculative Execution Barriers:} Speculative execution barriers, such as LFENCE, force the completion of all prior instructions before allowing further speculation, thereby preventing attacks like Spectre and Meltdown from accessing sensitive data speculatively \cite{milburn-2022-arxiv, shen-2018-acm, kocher-2018-sp}. Hardware mechanisms such as Non-Data-Assisting Speculation (NDA) restrict speculative data propagation while maintaining performance \cite{weisse-2019-micro}. Execute on Clear (EoC) ensures that instructions with uncertain security implications are handled safely \cite{meng-2023-iccd}, while Speculative Taint Tracking (STT) marks data as tainted during speculative execution to prevent leakage \cite{yu-2019-micro}. Conditional speculation selectively blocks unsafe instructions, aiming to retain overall performance while enhancing security \cite{li-2019-hpca}.

\textbf{Branch Target Buffer (BTB) Protections:} BTB protections are critical for mitigating Spectre Variant 2, which targets branch mispredictions. Context-sensitive fencing flushes BTB entries at context switches, minimizing cross-process interference \cite{taram-2019-asplos}. Retpoline replaces indirect jumps with a return sequence, which avoids speculative execution of attacker-controlled instructions \cite{kadir-2019-iceee}. MicroCFI enforces control-flow integrity within the microarchitecture, reducing the chance of speculative code execution in unauthorized paths \cite{jang-2023-ieeeaccess}.

\textbf{Memory Fencing:} Memory fencing ensures correct ordering of memory operations, preventing speculative loads and stores from bypassing essential security checks. Instructions such as SFENCE.VMA in RISC-V finalize memory operations before any further speculative execution, mitigating vulnerabilities like Spectre and Meltdown \cite{weisse-2019-micro, meng-2023-iccd, chen-2022-aspdac}. These fencing techniques also support system reliability and data integrity, especially in high-assurance environments \cite{yu-2020-isca, yan-2018-micro}.

\textbf{Cache Partitioning:} Cache partitioning (also called way-partitioning) isolates cache lines among different security contexts, limiting cache-based side channels. DAWG (Dynamically Allocated Way Guard) achieves secure partitioning by segregating cache hits, misses, and updates across domains \cite{kiriansky-2018-micro}. Cyclone detects cache contention leaks based on cyclic interference patterns \cite{harris-2019-micro}, and CEASER-S reduces eviction-based attacks by randomizing cache line mappings, making it harder for attackers to create effective eviction sets \cite{qureshi-2019-isca}.

\textbf{Cache Timing Manipulation:} Attacks exploiting cache timing can be mitigated by altering or masking cache access patterns. Guard Cache introduces artificial cache hits and misses, which obscure timing channels with minimal performance overhead \cite{mosquera-2023-ieee}. STAR randomizes cache state changes to limit both speculative and non-speculative leaks \cite{hu-2023-arxiv}, while RaS randomizes cache line fills by inserting future-use lines, complicating attempts to track cache states \cite{hu-2023-arxiv-random}. ReViCe pairs a victim cache with jitter injection to restore speculatively loaded data upon mis-speculation, further hindering timing-based leakage \cite{kim-2020-secdev}.

\textbf{Randomized Cache Architectures:} Randomized eviction policies complicate the predictability of cache usage, impeding cache-centric side-channel attacks. CEASER-S and ScatterCache rely on encryption-based randomization, refreshing keys periodically to disrupt eviction set creation \cite{qureshi-2018-micro, song-2021-sp}. CEASER-S divides the cache into partitions, and indirection table-based methods map encrypted addresses through tables that add further complexity to cache manipulation \cite{ramkrishnan-2019-arxiv}. When combined with adaptive strategies, these randomized policies strengthen defenses against timing-based cache attacks \cite{vila-2018-sp}.

\textbf{Return Stack Buffer (RSB) Management:} Effective RSB management helps defend against SpectreRSB, which alters return addresses to initiate speculative execution of attacker-injected instructions. RSB stuffing preloads benign return addresses during context switches, preserving valid return predictions \cite{Korshunov-2017-SpectreRSB}. This measure applies to a range of architectures, including ARM, and mitigates ret2spec exploits that leverage RSB vulnerabilities \cite{taheri-2023-arxiv, maisuradze-2018-ccs}. Further improvements involve hardware-based return address validation, which checks return predictions before speculation proceeds \cite{shen-2019-arxiv}.

\textbf{Load and Store Hardening:} Load and store hardening prevents attacks like Load Value Injection (LVI) and Spectre-STL by enforcing checks on memory operations prior to speculation. Hardware modifications perform dependency checks to ensure speculative loads do not bypass related stores, while compiler-level transformations reorder instructions to preserve execution dependencies \cite{Kiriansky-2018-ArXiv, schaik-2019-sp, bulck-2020-sp}. The Store Vulnerability Window (SVW) reduces load re-execution, minimizing performance penalties \cite{roth-2005-isca}. InvisiSpec and Speculative Load Hardening (SLH) prevent improper speculative loads that might circumvent security validations \cite{yan-2018-micro, zhang-2022-iacr}.

\textbf{Data Execution Prevention (DEP):} DEP thwarts speculative code execution attacks by restricting execution permissions in non-executable memory areas. Coupled with Address Space Layout Randomization (ASLR), it complicates the prediction of executable code locations. While earlier DEP implementations (e.g., in Windows XP SP2) were vulnerable to return-oriented programming (ROP) \cite{stojanovski-2007-ares}, modern hardware-assisted defenses like Execute on Clear (EoC) reinforce DEP in conjunction with ASLR and control-flow integrity (CFI), thereby offering stronger protection against speculative attacks \cite{meng-2023-iccd, hu-2021-seed}.

\textbf{Software-Based Mitigation:} Compiler-level and runtime-level updates insert barriers and fences to confine speculative execution. Techniques like ASLR and Kernel Page Table Isolation (KPTI) rearrange memory layouts and isolate kernel memory from user processes, reducing exposure to speculative vulnerabilities \cite{rajasekaran-2020-acm, liu-2023-hasp}. Co-design frameworks such as SpecControl share branch dependency information with hardware to reduce performance overheads \cite{hajiabadi-2023-arxiv}, while frameworks like HyperRace shield SGX enclaves from transient execution exploits \cite{chen-2019-dsc}.

\textbf{Machine Learning Approaches:} Machine learning methods detect speculative execution attacks by analyzing performance metrics, correlation patterns, and other subtle signals in system behavior \cite{dutta-2019-acm}. Techniques such as STT mark data flows in speculative paths to prevent unauthorized propagation \cite{yu-2019-cacm}, and Specularizer monitors hardware performance counters for signs of malicious speculation \cite{wang-2021-sp}. Models like Non-Delay Assumption (NDA) break wrong-path dependencies to reduce speculative data leaks, offering an additional layer of defense \cite{xu-2017-icccn, weisse-2019-micro}.

\subsection{DPA Attacks}

\textbf{AES Algorithm Implementations:} DPA attacks on AES algorithm implementations are highly effective at extracting secret keys by analyzing power consumption variations during encryption operations. Simulated DPA on gate-level AES models has demonstrated the feasibility of retrieving cryptographic keys, confirming the vulnerability of such implementations \cite{smith2010methodology}. Hardware-based AES implementations on FPGAs are especially prone to these attacks, emphasizing the need for more robust countermeasures \cite{patel2012gpu}. Continued advancements in attack methodologies targeting hardware optimizations have further increased the efficiency of such attacks \cite{ambrose2008anatomy}.

\textbf{DES Algorithm Implementations:} DPA attacks targeting the DES algorithm have successfully recovered portions of the cryptographic key by analyzing power traces \cite{standaert2004fpga}. Techniques like signal companding help reduce the number of traces required for key extraction by mitigating noise interference \cite{ryoo2008performance}. These studies emphasize that hardware implementations, including FPGA-based designs, remain vulnerable and require enhanced defenses \cite{standaert2004fpga}.

\textbf{RSA Algorithm Implementations:} DPA attacks on RSA exploit power consumption variations during cryptographic operations to extract private keys \cite{li2012research}. Randomized masking has been proposed to counteract these vulnerabilities by reducing correlations between power consumption and the algorithm’s internal processes \cite{haitao2009differential}. Nonetheless, FPGA-based RSA systems remain susceptible, even when employing techniques like Montgomery multiplication \cite{bayam2008differential}, although shadow technology has shown promise in mitigating both DPA and Simple Power Analysis (SPA) threats \cite{kim2010correlation}.

\textbf{Elliptic Curve Cryptography (ECC) Implementations:} ECC implementations are also subject to DPA attacks that exploit power variations during scalar multiplication \cite{suresh2014dualfield}. Adversaries often target initial values in ECC computations to bypass standard countermeasures \cite{wang2009dparesistant}. Despite improvements in ECC security, statistical analysis of power traces still serves as an effective method for key extraction \cite{liang2007countermeasures}.

\textbf{SHA-3 Implementations:} DPA attacks on SHA-3 have revealed weaknesses, particularly in certain Message Authentication Code (MAC) configurations \cite{gauravaram2008dpa}. Differential fault analysis further demonstrates the susceptibility of SHA-3’s internal state to side-channel attacks \cite{bagheri2015dpa}. Even with sophisticated masking techniques, hardware implementations remain vulnerable \cite{chu2018dpa}.

\textbf{Stream Cipher Implementations:} Stream ciphers such as Trivium and Grain are not immune to DPA attacks. Optimized DPA on Trivium can extract full keys \cite{tenasanchez2015optimizedc}, while resynchronizing variants of Grain (Grain v1 and Grain-128) introduce additional attack surfaces \cite{chakraborty2015combined}. Implementing SABL (Sense Amplifier-Based Logic) on FPGAs has been suggested as a viable approach for enhancing DPA resistance \cite{atani2008dparesistiveb}.

\textbf{Blowfish, Twofish, and Serpent Implementations:} Blowfish and Twofish exhibit vulnerabilities to both DPA and SPA attacks. In particular, Twofish key recovery attacks exploit weaknesses in the key scheduling process \cite{ortiz2016simple}. Serpent has been shown to be susceptible to differential-linear cryptanalysis across up to 12 rounds of encryption \cite{dunkelman2008serpent}. Twofish also features key-dependent operations that can undermine the efficacy of certain DPA countermeasures \cite{rudinger2009twofish}.

\subsection{Mitigation of DPA Attacks}

\textbf{Hardware-Level Countermeasures:} To counteract DPA, various hardware-level modifications have been proposed. Dynamic Voltage and Frequency Scaling (DVFS) obscures power consumption patterns, thereby improving security \cite{guilley2006dvfs}. Additionally, hardware-based masking techniques, where intermediate cryptographic values are concealed with random masks, prevent attackers from correlating power consumption with sensitive data \cite{messerges1999masking}. Another effective strategy is dual-rail precharge logic (DPL), which balances power consumption during cryptographic operations to eliminate information leakage \cite{tiri2004dpl}. Noise generation circuits and nanoscale devices that obfuscate power profiles further enhance the security of hardware implementations \cite{khedkar2015profile, zhou2005noise}.

\textbf{Algorithmic Countermeasures:} Algorithmic techniques, such as masking, force attackers to perform more complex higher-order attacks, significantly increasing the difficulty of extracting useful data \cite{ding2014model}. Fresh re-keying schemes, which frequently update session keys, limit the amount of side-channel leakage over time, thereby mitigating attack risks \cite{dobraunig2017}. Higher-order threshold implementations randomize power consumption across different computational phases \cite{bilgin2014threshold}. Advanced logic designs, such as adiabatic dynamic differential logic (ADDL), further obfuscate power consumption patterns, making it more difficult for attackers to derive meaningful correlations \cite{morrison2015dynamic}.

\textbf{Randomization Techniques:} Randomization techniques introduce unpredictability into the execution of cryptographic operations, helping mitigate the effectiveness of DPA attacks. Varying the sequence of cryptographic instructions and using randomized initial points (RIP) in elliptic curve cryptography are effective methods for masking power consumption \cite{itoh2006, lu2010}. Similarly, randomizing the representation of private keys, particularly in ECC, significantly complicates key extraction \cite{ebeid2003}.

\textbf{Noise Injection:} Injecting random noise into the power supply of cryptographic devices is a common strategy for obscuring power consumption patterns. Correlated power noise generators have demonstrated robust protection in FPGA-based systems with minimal performance overhead \cite{kamoun2009correlated}. Randomized multitopology logic (RMTL) further secures cryptographic devices by dynamically altering power profiles during execution \cite{avital2015randomized}. Noise injection approaches have also been extended into wireless communication systems as a defense against eavesdropping \cite{he2017artificial}.

\textbf{Cryptographic Protocols:} Cryptographic protocols that incorporate zero-knowledge proofs or secure multiparty computation (MPC) inherently limit direct data exposure, making DPA attacks less effective \cite{mamiya2004, gennaro2012}. Homomorphic encryption allows computations on encrypted data without decryption, adding another layer of resilience against side-channel threats \cite{gentry2009}. Additionally, protocols such as Authenticated Encryption with Associated Data (AEAD) and key exchange methods like Diffie-Hellman mitigate both data leakage and active adversaries \cite{rogaway2011, diffie1976}.

\subsection{Memory Attacks}

\textbf{Cold Boot Attacks}: Cold boot attacks exploit the remanence effect in DRAM, allowing attackers to recover sensitive data even after the system loses power. Yitbarek et al. showed that DDR4 systems remain vulnerable despite memory scramblers, enabling AES key extraction \cite{yitbarek2017}. Riebler et al. leveraged FPGA-accelerated key search methods to reduce key reconstruction time, making cold boot attacks more practical \cite{riebler2013}. Seol et al. proposed Amnesiac DRAM, which erases memory contents upon physical separation or tampering, preventing data extraction \cite{seol2019}. Halderman et al. demonstrated that cooling extends memory remanence, underscoring the need for stronger defenses \cite{halderman2008}.

\textbf{Memory Remanence Attacks}: Memory remanence refers to the persistence of data in memory after power loss, posing significant security risks. Anagnostopoulos et al. revealed that low temperatures prolong data retention in SRAM PUFs, enabling accurate cryptographic key extraction \cite{anagnostopoulos2018}. Yu et al. introduced rapid erasure techniques for powered-off SRAM, effectively mitigating remanence attacks \cite{yu2009}. Skorobogatov showed that even flash memory remains vulnerable, as data can be recovered after erase operations \cite{skorobogatov2005}.

\textbf{Rowhammer Attacks}: Rowhammer attacks manipulate DRAM cells by repeatedly accessing a specific row, causing bit flips in adjacent rows. Veen et al. demonstrated the feasibility of rowhammer attacks on mobile platforms, indicating shortcomings in existing defenses \cite{veen2016}. Jiang et al. analyzed DDR3 and DDR4 DRAM vulnerabilities, revealing that rowhammer can bypass many hardware protections \cite{jiang2021}. Zhang et al. provided a retrospective analysis highlighting the need for robust strategies to address rowhammer across diverse systems \cite{zhang2022}.

\textbf{DMA Attacks}: Direct Memory Access (DMA) allows peripherals to access system memory directly, bypassing the CPU. Stewin et al. proposed a detection method that continuously monitors bus activity for suspicious DMA operations \cite{stewin2013}. Markuze et al. demonstrated that malicious peripherals can overwrite kernel data structures through DMA, effectively controlling the system \cite{markuze2021}. Morgan showed that it is possible to bypass IOMMU configurations, underscoring the severity of DMA attacks \cite{morgan2016}.

\textbf{Bus Snooping Attacks}: Bus snooping attacks intercept data transmissions between the CPU and memory. Moon et al. introduced Vigilare, a bus-snooping-based kernel integrity monitor capable of detecting transient kernel rootkit attacks \cite{moon2017}. Xu et al. proposed a technique that monitors phase shifts in digital waveforms to identify bus snooping \cite{xu2021}. Liu et al. demonstrated that cloud environments are similarly exposed and recommended Oblivious RAM (ORAM) for mitigating snooping risks \cite{liu2013}.

\textbf{ECC Attacks}: Error-Correcting Code (ECC) memory is designed to detect and correct errors, but it can still be exploited to reveal original data. Cojocar et al. showed that carefully orchestrated bit flips can circumvent ECC protections in rowhammer scenarios \cite{cojocar2019}. Yang et al. found that systems using combined Reed-Solomon and Hamming codes remain vulnerable to targeted manipulation of ECC logic \cite{yang2012}. Restifo et al. demonstrated how flaws in ECC encoding can produce incorrect corrections, offering further opportunities for data leakage \cite{restifo2017}.

\textbf{MMU Attacks}: Memory Management Unit (MMU) attacks exploit the virtual-to-physical address translation process to access protected memory. Gras et al. developed a cache-based EVICT+TIME attack that bypasses Address Space Layout Randomization (ASLR) by leveraging MMU operations \cite{gras2017}. Markuze et al. revealed sub-page vulnerabilities in IOMMU protections, enabling unauthorized memory access \cite{markuze2021}. Kim et al. discussed side-channel attacks exploiting MMU timing flaws, affecting both local and remote systems \cite{kim2023}.

\textbf{Thermal Imaging Attacks}: Thermal imaging attacks rely on residual heat patterns left by memory operations to infer stored data. Wodo et al. demonstrated how thermal imaging can reveal security codes from keypads \cite{wodo2016}. Salvi et al. used infrared imaging to detect hardware Trojans in chips by analyzing thermal signatures \cite{salvi2020}. Lohrke et al. employed Thermal Laser Stimulation (TLS) to extract cryptographic keys from FPGAs, even when powered off \cite{lohrke2018}.

\textbf{Timing Attacks}: Timing attacks exploit variations in execution time during cryptographic operations to recover sensitive information. Hirata et al. investigated RSA decryption timing variations stemming from Montgomery exponentiation, enabling private key extraction \cite{hirata2020}. Jiang et al. performed a GPU-based timing attack on AES, targeting memory bank conflicts to recover keys \cite{jiang2017}. Huang et al. analyzed cache-timing attacks on HQC encryption, demonstrating chosen-ciphertext methods for key recovery \cite{huang2023}.

	\subsection{Mitigation of Memory Attacks}
	\textbf{Hardware-Based Memory Encryption:} RAM encryption provides important protection for volatile memory. Intel’s Total Memory Encryption (TME) safeguards data at rest in main memory with minimal performance overhead \cite{mofrad2018tc}, while AMD’s Secure Memory Encryption (SME) applies per-page encryption to further strengthen security without incurring significant slowdowns \cite{gueron2016ieee}. In virtualized environments, AMD’s Secure Encrypted Virtualization (SEV) extends these protections to virtual machines, reducing the risk of hypervisor-level attacks \cite{li2022cloud}. Nevertheless, research by Bühren et al. indicates that encryption alone does not defend against fault injection attacks, highlighting the need for supplementary integrity checks \cite{buhren2016}. Inoue et al. introduced the Scalable Memory Encryption Engine (SIT), which enhances encryption throughput for large-scale systems \cite{inoue2022micro}.

	\textbf{Secure Key Management:} Proper management of cryptographic keys is essential for overall security. Hardware Security Modules (HSMs) securely store private keys in many cloud environments, as shown by Yu et al. \cite{yu2018}. Gopal et al. demonstrated that smaller organizations can adopt Trusted Platform Module (TPM)-based key management for a cost-effective alternative \cite{gopal2018}. Rigorous standards, including FIPS 140-2, guarantee secure key transport and management in cryptoprocessors, as noted by Rajitha et al. \cite{rajitha2016}. For embedded devices, Obermaier et al. proposed Physical Unclonable Function (PUF)-based key management, which derives keys without relying on non-volatile memory \cite{obermaier2018}.

	\textbf{Memory Integrity Verification:} Various cryptographic mechanisms, such as hash trees and Message Authentication Codes (MACs), can verify memory integrity. Hou et al. combined these structures in a hardware scheme that lowers verification overhead \cite{hou2004}. Fang et al. improved performance through optimized hash tree checks, using hot-windows to shorten verification paths \cite{fang2004}. Vig et al. presented dynamic skewed integrity trees that adapt to memory access patterns, thereby enhancing efficiency \cite{vig2018}. Shwartz et al. proposed Distributed Integrity Trees, which facilitate low-overhead memory verification in parallelized or distributed settings \cite{shwartz-2018-ieee}.

	\textbf{Side-Channel Attack Mitigations:} Several strategies address side-channel threats in memory systems. Lee et al. used static analysis to detect rowhammer by identifying characteristic access patterns, improving detection accuracy \cite{lee2021detection}. Cojocar et al. demonstrated that ECC memory remains susceptible to rowhammer, as carefully orchestrated bit flips can circumvent ECC mechanisms \cite{cojocar2019eccploit}. Deutsch et al. introduced DAGguise, which employs Directed Acyclic Request Graphs to minimize memory timing leaks while preserving performance \cite{deutsch2022dagguise}. Jiang et al. developed MemPoline, a system that disguises memory access patterns and protects CPUs and GPUs from memory-based side channels \cite{jiang2020mempoline}.

	\textbf{Randomized Memory Access Patterns:} Randomizing memory access sequences can mitigate timing and side-channel attacks by reducing an attacker’s ability to predict data flow. Liu et al. showed that oblivious shuffling and address space randomization hamper reverse engineering in deep neural networks without imposing substantial overhead \cite{liu2019}. Jiang et al. introduced MemPoline, which scatters sensitive data to obfuscate memory access patterns and improve efficiency \cite{jiang2020mempoline}. Qureshi et al. proposed CEASER, encrypting cache line addresses and randomizing their placement in the cache, thereby offering robust protection against side-channel analysis \cite{qureshi-2018-micro}. Kadam et al. developed RCoal, which randomizes coalescing in GPU memory accesses, significantly reducing exposure to timing-based attacks \cite{kadam2018}.

	\textbf{Memory Scrubbing:} Memory scrubbing detects and corrects errors in memory systems to maintain data integrity. Saleh et al. introduced Single-Error-Correction and Double-Error-Detection (SEC-DED) codes, a foundational method for improving resilience against bit flips \cite{saleh1990}. Reviriego et al. examined optimized scrubbing sequences that markedly extend the mean time to failure (MTTF) of memory subsystems \cite{reviriego2010}. Li et al. demonstrated how deterministic and probabilistic scrubbing can bolster reliability in FPGA-based designs, with varying scrubbing rates and access patterns \cite{li2013}. Tonfat et al. proposed energy-efficient scrubbing using frame-level redundancy scrubbing (FLR), which enhances error correction while minimizing power draw \cite{tonfat2015}.
	
\subsection{Secure Enclave Attacks}
Secure enclaves, including Intel's SGX, ARM's TrustZone, and Apple's Secure Enclave, are robust by design but still vulnerable to a variety of sophisticated attack vectors. Interface attacks leverage the interaction between enclaves and host applications, as demonstrated by the COIN model, which highlights how exposed interfaces can lead to control-flow hijacking and information leaks in SGX projects \cite{Khandaker2020d403}. Side-channel attacks exploit shared microarchitectural states, making enclave isolation on separate cores a crucial mitigation \cite{Maas2020c8ca}. Microarchitectural vulnerabilities, such as Spectre and Meltdown, allow attackers to manipulate branch predictions, leaking sensitive data \cite{Chen20182df8}. Controlled-channel attacks utilize page table manipulations to reconstruct enclave control flows, but solutions like Autarky provide full control over page faults to obscure memory access patterns \cite{Moghimi20204a5b, Orenbach2020f39f}. Furthermore, interrupt-based side-channel attacks expose the need for secure interrupt handling \cite{Piessens20193593}. Memory attacks, which exploit legacy memory segmentation and SSL/TLS side-channel traces, underscore the importance of strong memory isolation and constant-time programming \cite{Xiao20171181}. Malware attacks targeting enclaves can impersonate host applications, steal sensitive data, or encrypt it for ransom, as discussed in the enclave malware research \cite{Schwarz20193e78}. Additionally, power side-channel attacks like PLATYPUS leak secrets such as AES-NI keys through variations in power consumption, further complicated by Intel’s RAPL interface \cite{Lipp20218fb3}. Finally, the rise of machine learning-enhanced side-channel attacks emphasizes the increasing complexity of defending enclaves \cite{Amrouche2022a68e}.

\subsection{Mitigation of Secure Enclave Attacks}
To counter these threats, several strategies have emerged. Physical isolation of enclaves on separate processor cores is an effective defense against intra-core and inter-core side-channel attacks \cite{Maas2020c8ca}. Speculative execution control in processors like MI6 introduces a purge instruction that ensures memory hierarchy isolation during enclave operations \cite{Bourgeat2018}. Automated security policy enforcement utilizes enclaves to enforce application-specific information security policies, protecting against low-level attacks \cite{Gollamudi2016}. To address controlled-channel attacks, systems like T-SGX suppress page fault notifications using Intel’s TSX, effectively reducing attack surface \cite{Shih2017}, while SGX-LEGO randomizes memory access sequences, rendering fault information unusable \cite{Kim2019}. Interruptible enclave execution security has improved through techniques such as full abstraction, preventing side-channel vulnerabilities during processor feature additions \cite{Busi2021}. Data location randomization breaks correlations between adversarial observations and actual memory access, providing a strong automated defense without developer intervention \cite{Brasser2019}. Lastly, integrating machine learning-based defenses anticipates and counters evolving machine learning-enhanced side-channel attacks \cite{Amrouche2022a68e}.

\subsection{Secure Boot and Root of Trust Attacks}
BootHole revealed critical vulnerabilities in the GRUB2 bootloader, where attackers could modify its configuration file to execute malicious code, bypassing secure boot protections \cite{Jackson2020}. Similarly, S3Booster exploits UEFI firmware vulnerabilities by inserting persistent malware during the system’s S3 sleep state \cite{Bulygin2014}. Thunderstrike is another physical attack leveraging Thunderbolt’s DMA access to inject malware into system firmware, bypassing software security mechanisms \cite{Kasper2015}. SMM rootkits operate within System Management Mode (SMM), injecting undetectable rootkits with the highest privilege level, bypassing the operating system \cite{Tang2009}. Power side-channel attacks on secure boot, using tools like ChipWhisperer, can extract cryptographic keys through power analysis during the boot process, further demonstrating the need for robust obfuscation techniques \cite{OFlynn2014}.

\subsection{Mitigation of Secure Boot and Root of Trust Attacks}
Several strategies are employed to mitigate these vulnerabilities. Trusted Platform Modules (TPMs) and Hardware Security Modules (HSMs) provide tamper-resistant environments for securely storing cryptographic keys, ensuring boot process integrity. For example, a recent study proposes a TEE SoC architecture that isolates the root of trust, protecting against unauthorized access post-boot \cite{Hoang2022Trusted}. HSM integration in virtualized cloud environments, such as with SvTPM, strengthens isolation from potentially malicious tenants \cite{Wang2023SvTPM}. Firmware integrity can be maintained through cryptographic signature verification, as implemented in BootKeeper, which validates boot firmware, preventing unauthorized updates \cite{Chevalier2019BootKeeper}. Physical security measures such as tamper-evident seals mitigate attacks like Evil Maid and Thunderstrike, which target firmware vulnerabilities \cite{b349}. Finally, monitoring systems that detect abnormal voltage and frequency fluctuations defend against glitching attacks, ensuring system integrity during the boot process \cite{Xie2020, Zhou2019}. Speculative execution controls are also critical in mitigating attacks like Spectre Boot, where speculative execution paths are exploited during boot to bypass security checks \cite{Kocher2019}.

\subsection{Fault Injection Attacks}
Fault injection attacks pose a significant threat to cryptographic systems by intentionally inducing hardware errors to manipulate the behavior of cryptographic operations and extract sensitive information. Common techniques include voltage glitching, where attackers momentarily reduce the power supply to cause timing errors; laser-induced faults, which introduce precise disruptions in circuits through targeted laser beams; and clock glitching, which manipulates clock signals to desynchronize critical operations like AES or RSA \cite{b357, b358, b359}. These methods enable attackers to perform differential fault analysis (DFA), using erroneous outputs to infer cryptographic keys. Additional methods, such as electromagnetic fault injection (EMFI) and thermal fault injection, introduce electromagnetic pulses or localized heating to disrupt device operations and reveal cryptographic secrets \cite{b361, b360}. The Rowhammer attack takes advantage of memory vulnerabilities in DRAM by repeatedly accessing rows to induce bit flips in adjacent rows, potentially leaking sensitive data \cite{b346}. Other fault injection techniques include power supply noise injection, infrared fault injection, and acoustic fault injection, where environmental disturbances induce faults, compromising cryptographic processes and leaking sensitive data \cite{b382, b370, b380}. These diverse techniques emphasize the growing need for robust countermeasures to safeguard cryptographic hardware from fault-based attacks \cite{b369}.

\subsection{Mitigation of Fault Injection Attacks}
Defending against fault injection attacks requires a multi-layered approach to enhance the robustness of cryptographic systems against various fault-based threats. To counter voltage and clock glitching, systems incorporate redundancy in critical operations and employ error detection and correction codes (EDC and ECC) to detect and correct faults in real-time, ensuring cryptographic computations remain secure despite glitches \cite{kim_2007_a1, mangard_2004_b2}. For laser fault injection attacks, physical defenses like laser-absorbing shielding materials are used, alongside logical techniques such as redundant computations and random delays, which prevent attackers from precisely targeting cryptographic operations \cite{boneh_1997_c3, genkin_2015_d4}. Thermal fault injection defenses monitor and control the temperature of critical components using thermal sensors and robust thermal management systems, which detect abnormal fluctuations and prevent fault induction \cite{barenghi_2012_e5, oswald_2018_f6}. Electromagnetic Fault Injection (EMFI) is mitigated through electromagnetic shielding, filtering techniques, and spread-spectrum clocking, which complicate the process of inducing faults via electromagnetic pulses \cite{skorobogatov_2002_g7, daemen_2002_h8}. Mitigations for Rowhammer attacks involve both hardware and software solutions, including error-correcting codes (ECC) and operating system protections that isolate sensitive memory regions to prevent data leakage from adjacent rows \cite{b346}. For attacks like Plundervolt, which exploit software-controlled undervolting, hardware locks are used to prevent unauthorized voltage adjustments, while real-time monitoring systems detect and block malicious undervolting attempts \cite{b370}. Comprehensive fault injection mitigation strategies combine physical shielding, redundancy, error detection, and control-flow hardening, alongside automated vulnerability assessments to ensure robust protection against increasingly sophisticated fault-based attacks \cite{nasahl_2022_o15, guilley_2010_ad30, delarea_2022_ae31}.
	
\section{A Focus on RISC-V}

RISC-V's open and modular architecture has established it as a pivotal platform for hardware security research, enabling an unprecedented level of scrutiny into vulnerabilities and fostering the development of custom defense mechanisms directly at the architectural level. Unlike proprietary systems, the transparency of RISC-V provides unmatched flexibility for designing, implementing, and rigorously testing innovative security features. This applies across a spectrum of applications, from addressing speculative execution vulnerabilities to fortifying cryptographic processes. The openness of the RISC-V ecosystem also encourages robust collaboration between academia and industry, significantly accelerating the translation of critical security advancements into practical, deployable technologies. Consequently, RISC-V has emerged as a vital testing ground for pioneering research, offering a reproducible framework that goes beyond the limitations of traditional proprietary platforms. This unique capability enables the design of robust, adaptable solutions to address the dynamic and increasingly complex threats in modern computing environments. Given RISC-V's exceptional capacity for driving security innovation, the subsequent section will delve into a detailed analysis of specific attacks and propose tailored mitigation strategies designed for RISC-V architectures.

\subsection{Speculative Execution}

RISC-V Spectre Mitigation \cite{Bualucea_Irofti_2022_random} focuses on software-based techniques to address Spectre-BTI and Spectre-RSB vulnerabilities in RISC-V processors. Like their x86 counterparts, RISC-V processors are vulnerable to Spectre attacks, leading to the development of low-overhead mitigations integrated into the LLVM toolchain. These defenses, inspired by x86 strategies, enhance security without imposing significant performance penalties. Publicly available programs and data ensure transparency and facilitate the continuous refinement of these defensive measures.

SpecTerminator \cite{Jin_He_Qiang_2022_random} introduces a defense framework that leverages optimized hardware taint tracking and instruction masking to classify sensitive instructions. By managing execution characteristics, speculative side channels are effectively mitigated. Evaluations on an FPGA-accelerated simulation platform reveal overheads of 2.6\% for memory hierarchy side channels and 6.0\% for all known Spectre variants. This approach balances robust protection against speculative execution vulnerabilities with minimal performance degradation.

The MINOTAuR Core \cite{Gruin_Carle_Rochange_Cassé_Sainrat_2023_random} offers an open-source RISC-V processor design that integrates speculative execution while maintaining timing predictability. The architecture incorporates features such as a return address stack and advanced cache replacement policies, achieving performance comparable to the Ariane core. This pipeline design demonstrates feasibility in combining high performance with enhanced security against timing anomalies caused by speculative execution.

Le et al. \cite{Le_Hoang_Dao_Tsukamoto_Suzaki_Pham_2021_random} propose a real-time detection mechanism that utilizes hardware performance counters and neural networks to identify cache side-channel attacks on RISC-V processors. With a multi-layer perceptron achieving over 99\% detection accuracy and negligible performance impact, this method continuously monitors CPU cache behavior to provide robust defenses against Spectre attacks in practical systems.

BasicBlocker \cite{Thoma_Feldtkeller_Krausz_Güneysu_Bernstein_2020_random} modifies the RISC-V ISA to mitigate speculative execution vulnerabilities by reducing performance gains from speculation. The introduction of BBRISC-V, evaluated on a soft core and compiler, effectively neutralizes speculative execution attacks while maintaining acceptable performance. This simplified ISA redesign facilitates easier security analysis and bolsters resistance to speculative vulnerabilities.

SpectreCheck \cite{Gu_Chen_Wang_Xie_2020_random} implements a proactive detection mechanism for speculative execution side channels during early processor design. Using the RISC-V BOOM processor as a test case, it generates branch directions to guide predictors and expose side channels for analysis. This method emphasizes early identification of vulnerabilities, reducing the potential for future exploits and strengthening processor security in the development stage.

The ProSpeCT Model \cite{Daniel_Bognar_Noorman_Bardin_Rezk_Piessens_2023_random} formalizes secure speculation using constant-time policies within a processor model. By tracking secrets through the pipeline, it ensures they do not influence the microarchitectural state during speculation. ProSpeCT defends against all known Spectre variants and demonstrates minimal hardware and performance costs in a prototype implementation on a RISC-V processor, proving that secure speculation is achievable with negligible overhead.

\subsection{Power Analysis}

Researchers \cite{b397} demonstrated the feasibility of key extraction during AES encryption on RISC-V processors using Correlation Power Analysis (CPA). By correlating detailed power traces with internal computations, they highlighted the critical need for hardware-software co-design strategies to protect cryptographic operations in RISC-V Systems-on-Chip (SoCs).

Werner et al. \cite{b398} systematically examined several physical attack vectors, including power analysis. Their findings revealed that cryptographic routines inadvertently leak sensitive data through power patterns. Proposed countermeasures ranged from balanced logic styles at the design level to software-based obfuscation techniques for intermediate values.

Ahmadi et al. \cite{b399} conducted a comprehensive review of side-channel attacks on RISC-V processors, particularly power analysis. They identified gaps in existing countermeasures and advocated for integrating masking schemes, randomized clocking, and dynamic power management into a unified defense framework. Additionally, they suggested that machine learning techniques could enhance automated detection of side-channel attacks.

Mulder et al. \cite{b400} introduced masking techniques directly embedded in RISC-V architectures to counter power analysis attacks. By obscuring key-dependent power signatures at the hardware level, this approach reduced reliance on software-only defenses, mitigating information leakage during critical computations.

Random Dynamic Frequency Scaling (RDFS), proposed in \cite{b401}, disrupts CPA by making operating frequencies unpredictable during cryptographic operations. This unpredictability hampers correlation efforts by attackers, while empirical results showed minimal performance trade-offs alongside enhanced resistance to power analysis.

Le et al. \cite{b402} presented a real-time detection system combining cache activity monitoring and power analysis using neural networks. Their multi-layer perceptron model detects anomalies in near real time, enabling rapid responses to thwart ongoing side-channel exploits and prevent data compromise.

Tran et al. \cite{b403} explored the impact of microarchitectural events on power traces. By identifying events with the most distinctive signatures, they proposed design modifications to minimize information leakage, providing a practical blueprint for more secure RISC-V SoC designs.

Abella et al. \cite{b404} offered a holistic approach to addressing power analysis vulnerabilities, integrating clock randomization, secure boot protocols, and continuous fault-injection monitoring. Their findings underscored the importance of layered security strategies for comprehensive protection against side-channel attacks.

Stangherlin and Sachdev \cite{b405} implemented Boolean masking in a secure RISC-V microprocessor. By masking sensitive variables, they disrupted the correlation between power traces and cryptographic computations, significantly complicating key extraction attempts.

Schmidt et al. \cite{b406} tackled power analysis in vector processors by dynamically regulating power allocation during cryptographic routines. This fine-grained power management approach minimized data-dependent power patterns, achieving enhanced security while maintaining high performance.

\subsection{Memory Encryption}

Cilardo et al. \cite{b407} proposed a ChaCha-based memory encryption solution designed for Field-Programmable Gate Arrays (FPGAs). By optimizing resource utilization, their approach provides robust data security with minimal performance overhead. Experimental results on FPGA platforms demonstrated encryption overheads comparable to existing state-of-the-art techniques, validating the practicality of ChaCha-based encryption in resource-constrained scenarios.

Yin et al. \cite{b408} introduced a high-performance memory encryption engine (HPME) for Trusted Execution Environments (TEEs) in RISC-V Systems on Chips (SoCs). Featuring reduced cryptographic cycle counts and built-in authentication mechanisms, HPME ensures comprehensive data protection for Keystone-based TEEs. The efficient design achieves critical security guarantees while maintaining high system performance.

Werner et al. \cite{b409} enhanced defenses against physical tampering and fault injection by incorporating a pipeline stage dedicated to encrypting and decrypting instruction streams. This novel method uses memory encryption to secure instruction-level operations on RISC-V platforms, raising the difficulty for adversaries attempting to access or manipulate sensitive code.

SMARTS (Secure Memory Assurance of RISC-V Trusted SoC) \cite{b410} targets low-power and performance-sensitive applications by selectively encrypting critical memory regions. This targeted encryption minimizes both storage and performance overheads, while maintaining the confidentiality and integrity of sensitive data in trusted SoCs.

Lu et al. \cite{b411} developed a programmable PCIe encryption system for RISC-V, achieving throughput rates of up to 18.8 Gbps. This design showcases RISC-V’s capability to meet large-scale encryption requirements, providing high-speed data protection without compromising system efficiency.

Stangherlin et al. \cite{b405} introduced a secure microprocessor architecture incorporating Boolean masking and clock randomization. While primarily aimed at mitigating side-channel attacks, these hardware-level defenses simultaneously enhance memory security, achieving a resilient architecture with minimal performance trade-offs.

Stoffelen \cite{b413} outlined assembly-level optimizations for cryptographic primitives on the RISC-V instruction set. These low-level improvements enhance the performance of public-key algorithms, laying the groundwork for more efficient memory encryption techniques that rely on optimized cryptographic kernels.

RISE (RISC-V SoC for En/Decryption Acceleration) \cite{b415} employs a pseudorandom number generator and data-level parallelism to implement homomorphic encryption for edge devices. This approach reduces energy consumption while maintaining high throughput, making RISE a strong candidate for secure computing environments where memory encryption and decryption are critical.

\subsection{Cryptographic Instruction Set Architecture (ISA)}

Marshall et al. \cite{b416} introduced Scalar Advanced Encryption Standard (AES) Instruction Set Extensions (ISEs) for RISC-V, supporting both 32-bit and 64-bit cores. By integrating specialized opcodes, their design achieves significant speedups over software-only AES implementations while keeping hardware overhead minimal.

Pan et al. \cite{b414} proposed a lightweight AES coprocessor leveraging RISC-V custom instructions. This design delivers a 37.9\% performance gain in encryption tasks, making it particularly advantageous for IoT systems requiring efficient parallel data handling. The work demonstrates how tailored instruction support can significantly enhance cryptographic throughput in resource-constrained environments.

Marshall et al. \cite{b417} explored the Draft RISC-V Scalar Cryptography Extensions, reporting considerable efficiency improvements in 32-bit implementations. These findings provide valuable insights for standardizing extensions to streamline cryptographic capabilities across diverse RISC-V platforms, optimizing security-critical operations.

Gewehr et al. \cite{b418} focused on instruction set extensions for embedded systems running algorithms like AES-128, AES-256, SHA-256, and SHA-512. Evaluations on the Ibex core showed significant reductions in clock cycles, memory usage, and energy consumption, highlighting the practicality of deploying optimized cryptographic instructions in resource-limited settings.

Sud et al. \cite{b419} demonstrated the efficacy of bit manipulation instructions in accelerating cryptographic computations, achieving a 38\% reduction in clock cycles for lightweight block ciphers like LEA, SIMON, and SPECK. Similarly, Babu et al. \cite{b420} utilized the RISC-V “B” extension to realize a 28\% performance boost and a 20\% reduction in memory usage for cryptographic workloads. These findings emphasize the critical role of efficient bit manipulation in enhancing RISC-V-based cryptographic solutions.

Koppelmann et al. \cite{b422} expanded on the utility of bit manipulation instructions, showing their effectiveness in improving runtime efficiency and lowering power consumption during cryptographic operations, all without affecting pipeline execution. This low-overhead approach provides a pathway for integrating security-driven optimizations into future RISC-V architectures.

Saarinen et al. \cite{b423} introduced a modular Entropy Source (ES) that decouples entropy generation from deterministic random bit generators (DRBGs). This design simplifies implementation and aligns with cryptographic standards, enhancing the robustness of RISC-V microcontrollers. Such innovations address the essential requirement for reliable randomness in secure communication protocols.

Fritzmann et al. \cite{b425} incorporated post-quantum cryptography (PQC) accelerators into RISC-V, introducing 29 new instructions optimized for schemes like NewHope, Kyber, and Saber. The work achieved significant gains in performance and energy efficiency, illustrating RISC-V’s capacity to handle next-generation cryptographic demands and establishing it as a leading choice for PQC deployments.

\section{Conclusion}

In an era of rapid technological advancements and widespread adoption of cloud services, IoT devices, and edge computing platforms, the security of underlying hardware has never been more critical. The complexity and openness of modern computing architectures as exemplified by RISC-V’s flexibility offer unparalleled opportunities for innovation but also introduce new avenues for malicious actors to exploit. Accordingly, a holistic understanding of hardware vulnerabilities and the development of robust countermeasures are paramount to safeguard confidentiality, integrity, and availability of sensitive data. This review aims to highlight the pressing need for concerted efforts by both academia and industry to anticipate emerging threats, foster collaboration, and engineer secure hardware architectures capable of withstanding increasingly sophisticated attacks.

In summary, this paper has provided a detailed analysis of modern hardware security challenges and their implications for contemporary and emerging computer architectures, with a special focus on RISC-V. A close examination of the explored attack vectors and mitigation strategies highlights that hardware security is a continuously evolving field, requiring ongoing research and development to address increasingly sophisticated threats. Open-source architectures, such as RISC-V, have introduced flexibility and innovation. However, the openness and configurability that benefit development also expose these architectures to various attacks, including side-channel attacks, hardware fault injections, and speculative execution vulnerabilities.

Advanced techniques like hardware-based isolation, secure microcode updates, speculative execution controls, cache partitioning, randomized execution, and enhanced error detection and correction mechanisms are essential for mitigating these vulnerabilities. While speculative execution improves performance, it has also been exploited, as demonstrated by vulnerabilities like Spectre and Meltdown. Persistent threats from cache side-channel attacks, including time-driven, trace-driven, and access-driven methods, require advanced countermeasures such as hardware-assisted buffer protection and control-flow integrity mechanisms. Power side-channel attacks, such as Simple Power Analysis (SPA) and Differential Power Analysis (DPA), exploit power consumption variations to extract sensitive information. Electromagnetic analysis (EMA) uses emissions from devices to recover cryptographic keys. Robust countermeasures, such as masking and frequent rekeying, are necessary to mitigate these risks.

Memory encryption, particularly selective and granular strategies, is critical for protecting sensitive data. Secure boot and root of trust mechanisms ensure devices boot with only trusted software, establishing a secure foundation. Integrating cryptographic operations directly into the processor's instruction set significantly improves both the performance and security of cryptographic computations, as demonstrated by technologies like Intel AES-NI, ARM Cryptographic Extensions, and RISC-V Cryptographic Extensions. Secure enclaves offer isolated execution environments that protect sensitive computations, and Physical Unclonable Functions (PUFs) leverage manufacturing variations to create unique device fingerprints for secure authentication and key generation.

This review underscores the importance of continued research and collaboration in hardware security. As technology advances, new vulnerabilities will emerge, necessitating innovative security measures. Therefore, research in this field should focus on strengthening hardware-based isolation techniques, developing secure microcode updates, improving speculative execution controls and cache partitioning strategies to mitigate side-channel attacks, exploring emerging cryptographic ISAs, and expanding the use of secure enclaves and PUFs to enhance overall system security.

Ensuring the security of modern and emerging computer architectures is an ongoing challenge. This paper contributes to the understanding of hardware security by identifying key vulnerabilities and proposing effective countermeasures. The findings emphasize the need for a proactive approach to hardware security, advocating for the integration of comprehensive security features in future architectures to counter evolving threats. Continued innovation and vigilance are essential to preserving the integrity and confidentiality of sensitive data in an increasingly interconnected world.

    \bibliographystyle{IEEEtran}
    \bibliography{fp}
    
    \end{document}